\documentclass[aps,prl,twocolumn, showpacs,
  superscriptaddress]{revtex4}

\usepackage{graphicx} \usepackage{color} \usepackage{natbib}

\begin{document}

   \title{Mass ejection by strange star mergers and observational
     implications}

\author{A.~Bauswein} \affiliation{Max-Planck-Institut f\"ur
  Astrophysik, Karl-Schwarzschild-Str.~1, D-85748 Garching, Germany}
\author{H.-T.~Janka} \affiliation{Max-Planck-Institut f\"ur
  Astrophysik, Karl-Schwarzschild-Str.~1, D-85748 Garching, Germany}
\author{R.~Oechslin} \affiliation{Max-Planck-Institut f\"ur
  Astrophysik, Karl-Schwarzschild-Str.~1, D-85748 Garching, Germany}

\author{G.~Pagliara} \affiliation{Institut~f\"ur~Theoretische~Physik,
  Universit\"at~Heidelberg, Philosophenweg~16, D-69120~Heidelberg,
  Germany} 
\author{I.~Sagert} \affiliation{Institut~f\"ur~Theoretische~Physik, Goethe~Universit\"at, Max-von-Laue Str.~1, D-60438 Frankfurt, Germany}

\author{J.~Schaffner-Bielich}
\affiliation{Institut~f\"ur~Theoretische~Physik, Universit\"at~Heidelberg, Philosophenweg~16, D-69120~Heidelberg, Germany}

\author{M.~M.~Hohle} \affiliation{Astrophysikalisches Institut und 
 Universit\"ats-Sternwarte, Schillerg\"asschen~2-3, D-07745 Jena, Germany}
\affiliation{Max-Planck-Institut f\"ur extraterrestrische Physik, Postfach 1312, D-85741 Garching, Germany}
\author{R.~Neuh\"auser} \affiliation{Astrophysikalisches Institut und
 Universit\"ats-Sternwarte, Schillerg\"asschen~2-3, D-07745 Jena, Germany}

	\date{\today} 

  \begin{abstract}
We determine the Galactic production rate of strangelets as a canonical input to calculations of the measurable cosmic ray flux of strangelets by performing simulations of strange star mergers and combining the results with recent estimates of stellar binary populations. We find that the flux depends sensitively on the bag constant of the
MIT bag model of QCD and disappears for high values of the bag
constant and thus more compact strange stars.  In the latter case strange stars could coexist with
ordinary neutron stars as they are not converted by the capture of cosmic ray
strangelets. An unambiguous detection of an ordinary neutron star
would then {\em not} rule out the strange matter hypothesis.
   \end{abstract}
   

   \maketitle

The strange matter hypothesis (SMH) considers the possibility that the
absolute ground state of matter might not be formed by iron but by  strange quark matter (SQM), a mixture of up, down and strange quarks \cite{PhysRevD.4.1601,PhysRevD.30.272}. If true, stable
objects consisting of this matter with baryon numbers from about
$10^2$ to $10^{57}$ might exist
\cite{1996csnp.book.....G,2005PrPNP..54..193W,2007ASSL..326.....H}. The latter end corresponds
to strange stars (SS) with a mass and radius comparable to that of
neutron stars (NS), where the upper mass limit is given by the
inevitable collapse to a black hole (BH) \cite{1986A&A...160..121H,1986ApJ...310..261A}. In contrast to NSs
these SSs are selfbound and do not have an overall inverse mass-radius
relation.

One of the astrophysical consequences of the SMH is the
possibility that collision events of two SSs lead to the ejection of
strangelets, small lumps of SQM
\cite{1988PhRvL..61.2909M,1991PhLB..264..143C}. These strangelets
would contribute to the cosmic ray flux. The pollution of the Galaxy with strangelets is speculated to convert all ordinary NSs to SSs. It was
argued that all compact stars have become SSs in this scenario because
already a tiny amount of strangelets is sufficient to trigger the
transformation \cite{1987PhLB..192...71O,1986ApJ...310..261A}. If this
sequence of arguments was true, the unambiguous observation of a NS
would rule out the SMH according to references
\cite{1988PhRvL..61.2909M,1991PhLB..264..143C}. 

Estimates of the strangelet flux use results of NS-NS merger simulations
\cite{2005PhRvD..71a4026M}, which are not necessarily reliable in the
case of SQM. No detailed simulations of SS coalescence have been conducted
so far, and the ejected mass is unknown.  Only Newtonian simulations
of SS-BH binaries, modeling the BH by a pseudo-relativistic potential
\cite{1980A&A....88...23P}
, were carried out by \cite{2002MNRAS.335L..29K}. It was found that from this kind of mergers
no matter is ejected.  In order to shed light on the merger process of
SS binaries we performed relativistic three-dimensional hydrodynamical
simulations of the coalescence.

Several current and upcoming 
experiments have the potential to detect signatures of SQM. For
instance SQM might be produced directly in the Large Hadron Collider
at CERN \cite{Greiner:1987tg,Spieles:1996is}.  But also cosmic ray
experiments like the Alpha Magnetic Spectrometer AMS-02 planned to be
installed on the International Space Station in 2010 are
designed to capture strangelets \cite{ams,2004JPhG...30S..51S}. In
addition, gravitational-wave detectors like LIGO and VIRGO might
identify characteristic signals from SS mergers and SS oscillations or
instabilities \cite{2007ASSL..326.....H}. Also indirectly, the observation of compact stars can
help to decide on the SMH especially by pinning down the mass-radius
relation \cite{2007ASSL..326.....H,2005PrPNP..54..193W,1996csnp.book.....G}. For a review
on additional SQM searches see
\cite{2005PrPNP..54..193W,2006JPhG...32S.251F}.

The expectation that
all NSs convert to SSs and that there is a measurable flux of
strangelets as cosmic rays, relies on the assumption that SS mergers or another source indeed eject SQM in a sufficient amount. Here we report that a pollution through SS 
mergers does not need not be present for all models describing absolutely stable
SQM. In fact, we find that the amount of ejected matter depends on the
so-called bag constant, which in the MIT bag model adopted here
represents the pressure of the non-perturbative QCD vacuum. Therefore
the determination of the mass flux of strangelets in cosmic rays could
help to constrain this unknown parameter, which in turn gives the
binding energy of SQM.  Complementary insights in the equation of
state (EoS) of SQM might come from the detection of gravitational
waves (GWs) from SS coalescences, if some specific features distinguished
them from those of merging NSs.

In order to describe the EoS of
absolutely stable SQM we employ the MIT bag model
\citep{PhysRevD.9.3471,Farhi:1984qu}. Within this model, quarks are
considered as a free or weakly interacting Fermi gas and the
non-perturbative QCD interaction is simulated by a finite pressure of
the vacuum, the bag constant $B$. The small current masses of up and
down quarks allow us to treat them as massless particles, whereas for
the strange quark we adopt the value of $m_s=100$ MeV
\cite{Eidelman:2004wy}.  For our study we consider free		 quarks, which
corresponds to a range of the bag constant of $57$ MeV/fm$^3 \lesssim
B \lesssim 84$ MeV/fm$^3$.  The lower limit of $B$ is given by the
fact that baryons do not convert spontaneously to a two flavor quark
phase, while the upper limit is determined by the requirement of
absolutely stable SQM (energy per baryon at zero pressure, $E/A$,
lower than the corresponding value of $930$ MeV for nuclear matter).
These limits can be altered for other choices of $m_s$ or by
considering the interactions among quarks. The values of $60$
MeV/fm$^3$ ($E/A=860$ MeV) and $80$ MeV/fm$^3$ ($E/A=921$ MeV) for the
bag constant have been chosen to represent the extreme cases of the
underlying microphysical model, which we refer to as MIT60 and MIT80.
This choice of parameters yields a maximum mass of bare cold SSs of
1.88 $M_{\odot}$ for MIT60 and 1.64 $M_{\odot}$ for MIT80 with
corresponding radii of 10.4 km and 9.0 km, respectively.  For given
mass the stellar radii for MIT80 are in general slightly smaller than
those for MIT60. A possible nuclear crust of SSs is neglected
because of its small mass ($\sim $10$^{-5}M_{\odot}$), which makes it irrelevant for the dynamics of the system.

We performed SS merger simulations with the code described in
\cite{2007A&A...467..395O}. A three-dimensional relativistic smoothed
particle hydrodynamics scheme (SPH) is combined with an approximate
treatment of general relativity (GR) employing the conformal flatness
condition, supplemented by a method accounting for the backreaction of GW emission on the fluid \cite{2007A&A...467..395O}. Magnetic fields are not
included, because they can be considered as unimportant for the dynamical
behaviour as long as the initial field strength inside the compact
star is below $\sim$10$^{16}\,$G \cite{2008PhRvD..78b4012L}.

The models of
our simulations are chosen such that they cover the whole potential mass
range of compact star binaries.  The gravitational masses of the stars
vary between 0.9 $M_{\odot}$ and roughly the maximum SS mass for each
EoS. The simulations start after a relaxation phase from a
quasi-equilibrium orbit about two revolutions before the actual
merger. SQM has a shear viscosity comparable to that of nuclear
matter \cite{Haensel:1991pi} (and with color-superconducting phases even lower \cite{2000PhRvL..85...10M}.), which is expected to be too low to yield
tidally locked systems \cite{1992ApJ...400..175B}; therefore we consider
irrotational configurations. Thermal effects are also taken into
account as these were shown to be important for merger simulations
with nuclear EoSs, in particular when the ejection of mass is of 
interest \cite{2007A&A...467..395O}.

In total we discuss results of 29
simulations for MIT60 and 19 for MIT80 with a resolution of about
130,000 SPH particles. Using nonuniform particle masses we achieve a
mass resolution of roughly $10^{-5} M_{\odot}$ in these runs.  The
results were tested for convergence by additional simulations with
higher resolution.

There are two possible outcomes of SS mergers. For
relatively high masses of the binary components the merged object
collapses promptly to a BH shortly after the stars come in
contact. The forthcoming formation of a BH is indicated by a steep
decrease of the lapse function. Also the central density increases
within a fraction of the sonic timescale to values of twice the maximum
density of a single nonrotating SS and thus exceeds the
maximum density of stable, uniformly rotating ``supermassive'' SSs. If
the masses are lower, the merger remnant can be transiently supported
against collapse mainly by differential rotation. Such a
``hypermassive object'' (HMO) \cite{2000ApJ...528L..29B} emits a
characteristic GW signal, which is sensitive to the
total mass of the binary and the EoS (see \cite{2007PhRvL..99l1102O}
for NS mergers). This will be analyzed in a separate publication.
Since the system mass is much larger than the mass limit of
supermassive SSs for the given EoS, the remnant collapses to a BH
after the angular momentum has been redistributed
\cite{2000ApJ...528L..29B}.

\begin{figure}
\begin{center}
\includegraphics[width=8.5cm]{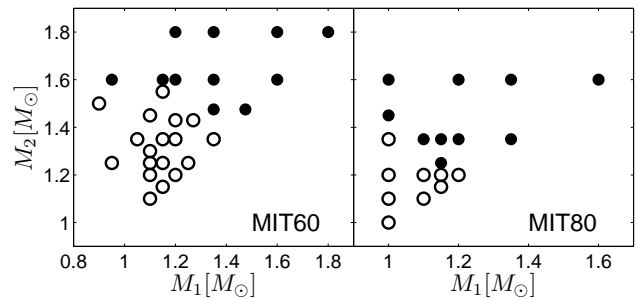}
\caption{Computed models for MIT60 and MIT80 in the $M_1$-$M_2$-plane
  of the gravitational masses of the SS binaries. Filled circles denote
  prompt collapse to a BH, while open circles indicate the formation
  of a HMO.}
\label{fig:models60}
\end{center}
\end{figure}

Figure~\ref{fig:models60} gives an
overview of the simulated binary mass configurations and their
outcome. Filled circles indicate prompt collapse to a BH while open
circles correspond to the formation of a HMO. $M_1$ and $M_2$ refer to
the gravitational masses of the SSs in isolation.

To estimate the amount of matter that becomes gravitationally unbound,
we use the criterion defined in \cite{2007A&A...467..395O}. It
considers the energy of a fluid particle in a comoving frame and
applies if pressure forces are small in comparison to gravitational
forces, which is well fulfilled for particles leaving the merger
site. In addition we
cross-checked these results by a simple criterion that monitors how
much matter expands away from the merger site. The ejecta estimates
agree within a factor of less than two. Ejecta from
HMOs originate from the tips of tidal tails that develop on timescales
longer than the timescale of prompt collapse to a BH. In the case of
such a prompt collapse no angular momentum can be redistributed from
the center to the outer parts of the merged object because the matter
in the inner part is swallowed quickly by the BH
\cite{2007A&A...467..395O}. 
Thus particles potentially forming an
accretion torus around the BH have no chance to end up in tidal tails
and to gain enough energy to become unbound.

\begin{figure}
\begin{center}
\includegraphics[width=8.5cm]{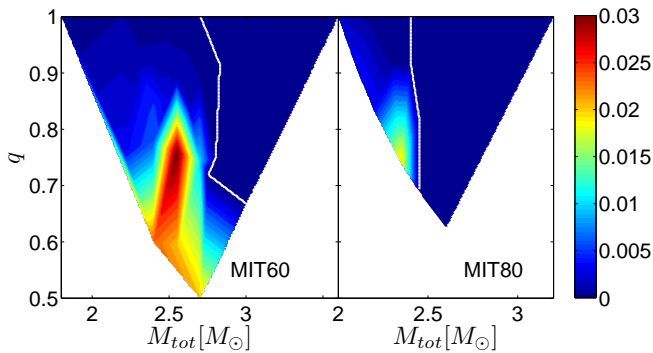}
\caption{Ejected mass per merger event, color-coded and measured in $M_{\odot}$, as
  function of the mass ratio $q=M_1/M_2$ and the total system mass
  $M_{tot}=M_1+M_2$ of the binary configurations for the MIT60 and
  MIT80 EoSs. The white line separates binary mergers with and without
  ejecta.}
\label{fig:ej60}
\end{center}
\end{figure}

Figure~\ref{fig:ej60} shows the estimated amount of unbound matter for
given mass ratios $q=M_1/M_2$ and total binary masses
$M_{tot}=M_1+M_2$ computed for MIT60 and MIT80 by means of our ejecta
criterion.  Above a certain $M_{tot}$ value we
cannot determine any amount of ejecta (see white lines in
Fig.~\ref{fig:ej60}). If $M_{tot}$ is below this limit, we obtain a
steep rise of the ejecta mass in a narrow region of the
$M_{tot}$-$q$-plane in both EoS cases for $q \lesssim 0.85$.  For
MIT60 the region where more than $0.01 M_{\odot}$ of matter become
unbound is located around a total mass of about 2.5 $M_{\odot}$.
For MIT80 significantly lower total masses are
required to obtain unbound matter and the ejected masses are lower
as well. This dependence on the bag constant originates from the fact
that MIT80 leads to more compact stars with correspondingly smaller
radii, which impedes the tidal disruption.

Since the ejected mass is
very low in comparison to the system mass, we found a dependence on
the chosen resolution and the initial setup of the SPH particles.  The
values of the ejected mass are uncertain within a factor of $\sim$2.
However, our conclusion that some configurations do not eject matter relies on
the occurrence of a prompt collapse to a BH. This is a safe result of
our simulations within the employed approximations. Therefore the
border between systems that eject matter and those that do not can be
considered as well determined (see Fig.~\ref{fig:ej60}). Only for
equal-mass binaries the borderline includes configurations that do not
collapse promptly and still do not eject matter, because such systems
do not form pronounced tidal arms (see \cite{2007A&A...467..395O} for 
NS mergers).

Population synthesis studies \cite{2008ApJ...680L.129B} provide probability
distributions of compact star binaries dependent on their system
parameters (e.g. $q$ and $M_{tot}$). Folding our results for the
ejecta masses with these probability distributions allows us to estimate
the ejected mass per merger event averaged over the whole
population. These numbers can be used to derive more accurately the
expected flux of strangelets in a detector like AMS-02
\cite{2005PhRvD..71a4026M}.  Assuming that the results of
\cite{2008ApJ...680L.129B} hold also for SSs and not only for NSs, we
compute for MIT60 a population-averaged ejecta mass of $8\cdot10^{-5}
M_{\odot}$. The uncertainties due to the limited resolution and the
criterion for determining ejecta masses can change this result up to a
factor of $\sim$4. For MIT80 we do not find any ejecta because
only configurations not present in the adopted population eject
matter.

For a rough assessment of the uncertainties
associated with the theoretical population synthesis studies we
employed a second data set based on observations of massive
progenitor stars in double systems \cite{hohle}. Using theoretical
results for the mass relation between NSs and progenitor stars \cite{2008ApJ...679..639Z}
and ignoring complications due to binary evolution effects, we derive a
probability distribution function of compact binaries.  Taking into
account uncertainties in the determination of stellar masses, we
obtain an average ejecta mass per event of 
$(1.4\,...\,2.8)\cdot 10^{-4}\,M_{\odot}$ for MIT60 and again a 
vanishing ejecta mass for MIT80.

The bag constant $B$ is the only parameter varied
between the EoSs and determines the mass-radius relation of SSs as the
crucial property
for the merger dynamics \cite{2007A&A...467..395O,2002PhRvL..89w1102F}. For intermediate values of $B$ we expect smaller ejecta masses than MIT60 but higher than MIT80. The borderline between models with and without
ejecta would then be shifted to an intermediate location as
well.

QCD perturbative corrections can be
absorbed in an effective bag constant that can be chosen to yield
mass-radius relations which agree well with the bag models we used \cite{Fraga:2001id}. Color superconductivity has only a small effect on the EoS \cite{1998PhLB..422..247A,Alford:2002rj}. However, quark interactions change the
$B$-window for absolutely stable SQM \cite{Farhi:1984qu,2001PhRvL..87q2003M,PhysRevD.66.074017,2007ASSL..326.....H}.

Our findings have important observational implications. 
The mass-radius relation (in our study determined by the bag constant) strongly affects the amount of matter ejected from SS mergers. Therefore a measured mass flux of
strangelets would constrain this relation if
quark star mergers were the main source of strangelets.
A relatively high flux would be an indicator for 
less compact SSs, while no or only a low flux
would only be consistent with more compact SSs. This would also put limits on the bag constant and so the binding energy of SQM.
 Assuming a Galactic merger rate of SS
binaries of $10^{-5}...10^{-4}\,\text{yr}^{-1}$
\cite{2008ApJ...680L.129B}, our population-averaged ejecta mass of
$\sim$10$^{-4} M_{\odot}$ for MIT60 yields a Galactic strangelet
production rate of $\dot{M}=10^{-9}...10^{-8}\,M_{\odot}\, 
\text{yr}^{-1}$. Since the flux of strangelets near Earth
depends linearly on $\dot{M}$, we derive a 10 to 100 times larger
value than in \cite{2005PhRvD..71a4026M}.

Even more relevant are
the consequences if there are no other production mechanisms of
strangelets. Our results for MIT80 imply that the
SMH cannot be ruled out but would only be compatible with compact SSs, if experiments like
AMS-02 could not find any evidence for a nonzero strangelet flux.  
A strangelet flux below a critical limit would mean that no or not all
NSs might have converted to SSs by capturing a strange nugget. SSs
might then still form by nucleation of SQM drops, e.g.\ during
stellar core collapse and explosion or by mass accretion of NSs in
binaries when the central conditions reach some critical threshold
for the phase transition to quark matter
\cite{1987PhLB..192...71O,1994PhRvD..49.2698O,2005PrPNP..54..193W,2007ASSL..326.....H}.
In this scenario there is a limiting mass above which SSs are formed
while NSs exist below (see also \cite{Bombaci:2004mt}). Thus in the
case of a large bag constant NSs and SSs could be in coexistence. In
the light of our simulations the unambiguous observation of a NS would
not rule out the SMH contrary to the suggestion in
\cite{1988PhRvL..61.2909M,1991PhLB..264..143C}.

We stress that these
conclusions from our simulations hold only if SS mergers are the only efficient sources of
strangelet ejection. In fact, several other suggestions have been
made, e.g.\ core-collapse supernova explosions
\cite{PhysRevD.57.5959} or the ejection by electric fields from the
surface of a SS \cite{2006PhRvD..74l7303C} if SQM nuggets were
embedded in the crust \cite{2006PhRvL..96d1101J}. Despite the
remaining uncertainties of our simulations like the approximate
treatment of GR, which also does not allow us to follow the formation
of the BH, the limited mass resolution, the simplified EoS, and the
omission of magnetic fields and a nuclear crust, we expect that a more
sophisticated approach will only yield quantitative shifts, changing the
exact values of the ejecta masses and possibly insignificantly moving the border
between configurations with and without ejecta.

Our results might also apply to other forms of selfbound matter like pion-condensed nucleon matter \cite{1990PhR...192..179M,2007ASSL..326.....H} provided the stellar properties are similar.

Future investigations, which should preferably be done in full GR, should
consider EoSs including quark interactions and
color superconductivity, or should use
descriptions beyond the MIT bag model. Also SS-BH mergers
should be reexamined in the GR framework. Finally, GW
signals from mergers besides the stellar cooling behaviour \cite{Yakovlev:2004iq,2005PrPNP..54..193W} may be promising means to
decide on the SMH or other forms of selfbound matter, once wave measurements will become possible.
\begin{acknowledgments}
This work was supported by DFG grants SFB/TR~7, SFB/TR~27, 
EXC~153, by the Heidelberg Graduate School of Fundamental Physics, by ESF/CompStar, by Helmholtz Alliance HA216/EMMI, by the Helmholtz Research School for Quark Matter Studies, and by computer time at LRZ Munich and RZG
Garching.
\end{acknowledgments}


\begin{thebibliography}{44}
\expandafter\ifx\csname natexlab\endcsname\relax\def\natexlab#1{#1}\fi
\expandafter\ifx\csname bibnamefont\endcsname\relax
  \def\bibnamefont#1{#1}\fi
\expandafter\ifx\csname bibfnamefont\endcsname\relax
  \def\bibfnamefont#1{#1}\fi
\expandafter\ifx\csname citenamefont\endcsname\relax
  \def\citenamefont#1{#1}\fi
\expandafter\ifx\csname url\endcsname\relax
  \def\url#1{\texttt{#1}}\fi
\expandafter\ifx\csname urlprefix\endcsname\relax\def\urlprefix{URL }\fi
\providecommand{\bibinfo}[2]{#2}
\providecommand{\eprint}[2][]{\url{#2}}

\bibitem[{\citenamefont{Bodmer}(1971)}]{PhysRevD.4.1601}
\bibinfo{author}{\bibfnamefont{A.~R.} \bibnamefont{Bodmer}},
  \bibinfo{journal}{Phys. Rev. D} \textbf{\bibinfo{volume}{4}},
  \bibinfo{pages}{1601} (\bibinfo{year}{1971}).

\bibitem[{\citenamefont{Witten}(1984)}]{PhysRevD.30.272}
\bibinfo{author}{\bibfnamefont{E.}~\bibnamefont{Witten}},
  \bibinfo{journal}{Phys. Rev. D} \textbf{\bibinfo{volume}{30}},
  \bibinfo{pages}{272} (\bibinfo{year}{1984}).

\bibitem[{\citenamefont{{Glendenning}}(1996)}]{1996csnp.book.....G}
\bibinfo{author}{\bibfnamefont{N.}~\bibnamefont{{Glendenning}}},
  \emph{\bibinfo{title}{{Compact Stars}}} (\bibinfo{publisher}{Springer-Verlag,
  New York}, \bibinfo{year}{1996}).

\bibitem[{\citenamefont{{Weber}}(2005)}]{2005PrPNP..54..193W}
\bibinfo{author}{\bibfnamefont{F.}~\bibnamefont{{Weber}}},
  \bibinfo{journal}{Progress in Particle and Nuclear Physics}
  \textbf{\bibinfo{volume}{54}}, \bibinfo{pages}{193} (\bibinfo{year}{2005}).

\bibitem[{\citenamefont{{Haensel} et~al.}(2007)\citenamefont{{Haensel},
  {Potekhin}, and {Yakovlev}}}]{2007ASSL..326.....H}
\bibinfo{author}{\bibfnamefont{P.}~\bibnamefont{{Haensel}}},
  \bibinfo{author}{\bibfnamefont{A.~Y.} \bibnamefont{{Potekhin}}},
  \bibnamefont{and} \bibinfo{author}{\bibfnamefont{D.~G.}
  \bibnamefont{{Yakovlev}}}, \emph{\bibinfo{title}{{Neutron Stars 1}}}
  (\bibinfo{publisher}{Springer-Verlag, New York}, \bibinfo{year}{2007}).

\bibitem[{\citenamefont{{Haensel} et~al.}(1986)\citenamefont{{Haensel},
  {Zdunik}, and {Schaefer}}}]{1986A&A...160..121H}
\bibinfo{author}{\bibfnamefont{P.}~\bibnamefont{{Haensel}}},
  \bibinfo{author}{\bibfnamefont{J.~L.} \bibnamefont{{Zdunik}}},
  \bibnamefont{and}
  \bibinfo{author}{\bibfnamefont{R.}~\bibnamefont{{Schaefer}}},
  \bibinfo{journal}{Astron. Astrophys.} \textbf{\bibinfo{volume}{160}},
  \bibinfo{pages}{121} (\bibinfo{year}{1986}).

\bibitem[{\citenamefont{{Alcock} et~al.}(1986)\citenamefont{{Alcock}, {Farhi},
  and {Olinto}}}]{1986ApJ...310..261A}
\bibinfo{author}{\bibfnamefont{C.}~\bibnamefont{{Alcock}}},
  \bibinfo{author}{\bibfnamefont{E.}~\bibnamefont{{Farhi}}}, \bibnamefont{and}
  \bibinfo{author}{\bibfnamefont{A.}~\bibnamefont{{Olinto}}},
  \bibinfo{journal}{Astrophys. J.} \textbf{\bibinfo{volume}{310}},
  \bibinfo{pages}{261} (\bibinfo{year}{1986}).

\bibitem[{\citenamefont{{Madsen}}(1988)}]{1988PhRvL..61.2909M}
\bibinfo{author}{\bibfnamefont{J.}~\bibnamefont{{Madsen}}},
  \bibinfo{journal}{Phys. Rev. Lett.} \textbf{\bibinfo{volume}{61}},
  \bibinfo{pages}{2909} (\bibinfo{year}{1988}).

\bibitem[{\citenamefont{{Caldwell} and {Friedman}}(1991)}]{1991PhLB..264..143C}
\bibinfo{author}{\bibfnamefont{R.~R.} \bibnamefont{{Caldwell}}}
  \bibnamefont{and} \bibinfo{author}{\bibfnamefont{J.~L.}
  \bibnamefont{{Friedman}}}, \bibinfo{journal}{Phys. Lett. B}
  \textbf{\bibinfo{volume}{264}}, \bibinfo{pages}{143} (\bibinfo{year}{1991}).

\bibitem[{\citenamefont{{Olinto}}(1987)}]{1987PhLB..192...71O}
\bibinfo{author}{\bibfnamefont{A.~V.} \bibnamefont{{Olinto}}},
  \bibinfo{journal}{Phys. Lett. B} \textbf{\bibinfo{volume}{192}},
  \bibinfo{pages}{71} (\bibinfo{year}{1987}).

\bibitem[{\citenamefont{{Madsen}}(2005)}]{2005PhRvD..71a4026M}
\bibinfo{author}{\bibfnamefont{J.}~\bibnamefont{{Madsen}}},
  \bibinfo{journal}{\prd} \textbf{\bibinfo{volume}{71}},
  \bibinfo{pages}{014026} (\bibinfo{year}{2005}).

\bibitem[{\citenamefont{{Paczy{\'n}ski} and
  {Wiita}}(1980)}]{1980A&A....88...23P}
\bibinfo{author}{\bibfnamefont{B.}~\bibnamefont{{Paczy{\'n}ski}}}
  \bibnamefont{and} \bibinfo{author}{\bibfnamefont{P.~J.}
  \bibnamefont{{Wiita}}}, \bibinfo{journal}{Astron. Astrophys.}
  \textbf{\bibinfo{volume}{88}}, \bibinfo{pages}{23} (\bibinfo{year}{1980}).

\bibitem[{\citenamefont{{Klu{\'z}niak} and {Lee}}(2002)}]{2002MNRAS.335L..29K}
\bibinfo{author}{\bibfnamefont{W.}~\bibnamefont{{Klu{\'z}niak}}}
  \bibnamefont{and} \bibinfo{author}{\bibfnamefont{W.~H.} \bibnamefont{{Lee}}},
  \bibinfo{journal}{Mon. Not. R. Astron. Soc.} \textbf{\bibinfo{volume}{335}},
  \bibinfo{pages}{L29} (\bibinfo{year}{2002}).

\bibitem[{\citenamefont{Greiner et~al.}(1987)\citenamefont{Greiner, Koch, and
  Stoecker}}]{Greiner:1987tg}
\bibinfo{author}{\bibfnamefont{C.}~\bibnamefont{Greiner}},
  \bibinfo{author}{\bibfnamefont{P.}~\bibnamefont{Koch}}, \bibnamefont{and}
  \bibinfo{author}{\bibfnamefont{H.}~\bibnamefont{Stoecker}},
  \bibinfo{journal}{Phys. Rev. Lett.} \textbf{\bibinfo{volume}{58}},
  \bibinfo{pages}{1825} (\bibinfo{year}{1987}).

\bibitem[{\citenamefont{Spieles et~al.}(1996)}]{Spieles:1996is}
\bibinfo{author}{\bibfnamefont{C.}~\bibnamefont{Spieles}} \bibnamefont{et~al.},
  \bibinfo{journal}{Phys. Rev. Lett.} \textbf{\bibinfo{volume}{76}},
  \bibinfo{pages}{1776} (\bibinfo{year}{1996}).

\bibitem[{\citenamefont{http://ams.cern.ch/}()}]{ams}
\bibinfo{author}{\bibnamefont{http://ams.cern.ch/}}.

\bibitem[{\citenamefont{{Sandweiss}}(2004)}]{2004JPhG...30S..51S}
\bibinfo{author}{\bibfnamefont{J.}~\bibnamefont{{Sandweiss}}},
  \bibinfo{journal}{Journal of Physics G} \textbf{\bibinfo{volume}{30}},
  \bibinfo{pages}{51} (\bibinfo{year}{2004}).

\bibitem[{\citenamefont{{Finch}}(2006)}]{2006JPhG...32S.251F}
\bibinfo{author}{\bibfnamefont{E.}~\bibnamefont{{Finch}}},
  \bibinfo{journal}{Journal of Physics G} \textbf{\bibinfo{volume}{32}},
  \bibinfo{pages}{251} (\bibinfo{year}{2006}).

\bibitem[{\citenamefont{Chodos et~al.}(1974)\citenamefont{Chodos, Jaffe,
  Johnson, Thorn, and Weisskopf}}]{PhysRevD.9.3471}
\bibinfo{author}{\bibfnamefont{A.}~\bibnamefont{Chodos}},
  \bibinfo{author}{\bibfnamefont{R.~L.} \bibnamefont{Jaffe}},
  \bibinfo{author}{\bibfnamefont{K.}~\bibnamefont{Johnson}},
  \bibinfo{author}{\bibfnamefont{C.~B.} \bibnamefont{Thorn}}, \bibnamefont{and}
  \bibinfo{author}{\bibfnamefont{V.~F.} \bibnamefont{Weisskopf}},
  \bibinfo{journal}{Phys. Rev. D} \textbf{\bibinfo{volume}{9}},
  \bibinfo{pages}{3471} (\bibinfo{year}{1974}).

\bibitem[{\citenamefont{Farhi and Jaffe}(1984)}]{Farhi:1984qu}
\bibinfo{author}{\bibfnamefont{E.}~\bibnamefont{Farhi}} \bibnamefont{and}
  \bibinfo{author}{\bibfnamefont{R.~L.} \bibnamefont{Jaffe}},
  \bibinfo{journal}{Phys. Rev.} \textbf{\bibinfo{volume}{D30}},
  \bibinfo{pages}{2379} (\bibinfo{year}{1984}).

\bibitem[{\citenamefont{Eidelman et~al.}(2004)}]{Eidelman:2004wy}
\bibinfo{author}{\bibfnamefont{S.}~\bibnamefont{Eidelman}} \bibnamefont{et~al.}
  (\bibinfo{collaboration}{Particle Data Group}), \bibinfo{journal}{Phys.
  Lett.} \textbf{\bibinfo{volume}{B592}}, \bibinfo{pages}{1}
  (\bibinfo{year}{2004}).

\bibitem[{\citenamefont{{Oechslin} et~al.}(2007)\citenamefont{{Oechslin},
  {Janka}, and {Marek}}}]{2007A&A...467..395O}
\bibinfo{author}{\bibfnamefont{R.}~\bibnamefont{{Oechslin}}},
  \bibinfo{author}{\bibfnamefont{H.-T.} \bibnamefont{{Janka}}},
  \bibnamefont{and} \bibinfo{author}{\bibfnamefont{A.}~\bibnamefont{{Marek}}},
  \bibinfo{journal}{Astron. Astrophys.} \textbf{\bibinfo{volume}{467}},
  \bibinfo{pages}{395} (\bibinfo{year}{2007}).

\bibitem[{\citenamefont{{Liu} et~al.}(2008)\citenamefont{{Liu}, {Shapiro},
  {Etienne}, and {Taniguchi}}}]{2008PhRvD..78b4012L}
\bibinfo{author}{\bibfnamefont{Y.~T.} \bibnamefont{{Liu}}},
  \bibinfo{author}{\bibfnamefont{S.~L.} \bibnamefont{{Shapiro}}},
  \bibinfo{author}{\bibfnamefont{Z.~B.} \bibnamefont{{Etienne}}},
  \bibnamefont{and}
  \bibinfo{author}{\bibfnamefont{K.}~\bibnamefont{{Taniguchi}}},
  \bibinfo{journal}{\prd} \textbf{\bibinfo{volume}{78}},
  \bibinfo{pages}{024012} (\bibinfo{year}{2008}).

\bibitem[{\citenamefont{Haensel}(1991)}]{Haensel:1991pi}
\bibinfo{author}{\bibfnamefont{P.}~\bibnamefont{Haensel}},
  \bibinfo{journal}{Nucl. Phys. Proc. Suppl.} \textbf{\bibinfo{volume}{24B}},
  \bibinfo{pages}{23} (\bibinfo{year}{1991}).

\bibitem[{\citenamefont{{Madsen}}(2000)}]{2000PhRvL..85...10M}
\bibinfo{author}{\bibfnamefont{J.}~\bibnamefont{{Madsen}}},
  \bibinfo{journal}{Physical Review Letters} \textbf{\bibinfo{volume}{85}},
  \bibinfo{pages}{10} (\bibinfo{year}{2000}).

\bibitem[{\citenamefont{{Bildsten} and {Cutler}}(1992)}]{1992ApJ...400..175B}
\bibinfo{author}{\bibfnamefont{L.}~\bibnamefont{{Bildsten}}} \bibnamefont{and}
  \bibinfo{author}{\bibfnamefont{C.}~\bibnamefont{{Cutler}}},
  \bibinfo{journal}{Astrophys. J.} \textbf{\bibinfo{volume}{400}},
  \bibinfo{pages}{175} (\bibinfo{year}{1992}).

\bibitem[{\citenamefont{{Baumgarte} et~al.}(2000)\citenamefont{{Baumgarte},
  {Shapiro}, and {Shibata}}}]{2000ApJ...528L..29B}
\bibinfo{author}{\bibfnamefont{T.~W.} \bibnamefont{{Baumgarte}}},
  \bibinfo{author}{\bibfnamefont{S.~L.} \bibnamefont{{Shapiro}}},
  \bibnamefont{and}
  \bibinfo{author}{\bibfnamefont{M.}~\bibnamefont{{Shibata}}},
  \bibinfo{journal}{Astrophys. J. Lett.} \textbf{\bibinfo{volume}{528}},
  \bibinfo{pages}{L29} (\bibinfo{year}{2000}).

\bibitem[{\citenamefont{{Oechslin} and {Janka}}(2007)}]{2007PhRvL..99l1102O}
\bibinfo{author}{\bibfnamefont{R.}~\bibnamefont{{Oechslin}}} \bibnamefont{and}
  \bibinfo{author}{\bibfnamefont{H.-T.} \bibnamefont{{Janka}}},
  \bibinfo{journal}{Phys. Rev. Lett.} \textbf{\bibinfo{volume}{99}},
  \bibinfo{pages}{121102} (\bibinfo{year}{2007}).

\bibitem[{\citenamefont{{Belczynski} et~al.}(2008)\citenamefont{{Belczynski},
  {O'Shaughnessy}, {Kalogera}, {Rasio}, {Taam}, and
  {Bulik}}}]{2008ApJ...680L.129B}
\bibinfo{author}{\bibfnamefont{K.}~\bibnamefont{{Belczynski}}},
  \bibinfo{author}{\bibfnamefont{R.}~\bibnamefont{{O'Shaughnessy}}},
  \bibinfo{author}{\bibfnamefont{V.}~\bibnamefont{{Kalogera}}},
  \bibinfo{author}{\bibfnamefont{F.}~\bibnamefont{{Rasio}}},
  \bibinfo{author}{\bibfnamefont{R.~E.} \bibnamefont{{Taam}}},
  \bibnamefont{and} \bibinfo{author}{\bibfnamefont{T.}~\bibnamefont{{Bulik}}},
  \bibinfo{journal}{Astrophys. J. Lett.} \textbf{\bibinfo{volume}{680}},
  \bibinfo{pages}{L129} (\bibinfo{year}{2008}).

\bibitem[{\citenamefont{{Hohle} et~al.}(2009)}]{hohle}
\bibinfo{author}{\bibfnamefont{M.~M.} \bibnamefont{{Hohle}}}
  \bibnamefont{et~al.}, \emph{\bibinfo{title}{in preparation}}
  (\bibinfo{year}{2009}).

\bibitem[{\citenamefont{{Zhang} et~al.}(2008)\citenamefont{{Zhang}, {Woosley},
  and {Heger}}}]{2008ApJ...679..639Z}
\bibinfo{author}{\bibfnamefont{W.}~\bibnamefont{{Zhang}}},
  \bibinfo{author}{\bibfnamefont{S.~E.} \bibnamefont{{Woosley}}},
  \bibnamefont{and} \bibinfo{author}{\bibfnamefont{A.}~\bibnamefont{{Heger}}},
  \bibinfo{journal}{Astrophys. J.} \textbf{\bibinfo{volume}{679}},
  \bibinfo{pages}{639} (\bibinfo{year}{2008}).

\bibitem[{\citenamefont{{Faber} et~al.}(2002)\citenamefont{{Faber},
  {Grandcl{\'e}ment}, {Rasio}, and {Taniguchi}}}]{2002PhRvL..89w1102F}
\bibinfo{author}{\bibfnamefont{J.~A.} \bibnamefont{{Faber}}},
  \bibinfo{author}{\bibfnamefont{P.}~\bibnamefont{{Grandcl{\'e}ment}}},
  \bibinfo{author}{\bibfnamefont{F.~A.} \bibnamefont{{Rasio}}},
  \bibnamefont{and}
  \bibinfo{author}{\bibfnamefont{K.}~\bibnamefont{{Taniguchi}}},
  \bibinfo{journal}{Phys. Rev. Lett.} \textbf{\bibinfo{volume}{89}},
  \bibinfo{pages}{231102} (\bibinfo{year}{2002}).

\bibitem[{\citenamefont{Fraga et~al.}(2001)\citenamefont{Fraga, Pisarski, and
  Schaffner-Bielich}}]{Fraga:2001id}
\bibinfo{author}{\bibfnamefont{E.~S.} \bibnamefont{Fraga}},
  \bibinfo{author}{\bibfnamefont{R.~D.} \bibnamefont{Pisarski}},
  \bibnamefont{and}
  \bibinfo{author}{\bibfnamefont{J.}~\bibnamefont{Schaffner-Bielich}},
  \bibinfo{journal}{Phys. Rev.} \textbf{\bibinfo{volume}{D63}},
  \bibinfo{pages}{121702} (\bibinfo{year}{2001}).

\bibitem[{\citenamefont{{Alford} et~al.}(1998)\citenamefont{{Alford},
  {Rajagopal}, and {Wilczek}}}]{1998PhLB..422..247A}
\bibinfo{author}{\bibfnamefont{M.}~\bibnamefont{{Alford}}},
  \bibinfo{author}{\bibfnamefont{K.}~\bibnamefont{{Rajagopal}}},
  \bibnamefont{and}
  \bibinfo{author}{\bibfnamefont{F.}~\bibnamefont{{Wilczek}}},
  \bibinfo{journal}{Phys. Lett. B} \textbf{\bibinfo{volume}{422}},
  \bibinfo{pages}{247} (\bibinfo{year}{1998}).

\bibitem[{\citenamefont{Alford and Reddy}(2003)}]{Alford:2002rj}
\bibinfo{author}{\bibfnamefont{M.}~\bibnamefont{Alford}} \bibnamefont{and}
  \bibinfo{author}{\bibfnamefont{S.}~\bibnamefont{Reddy}},
  \bibinfo{journal}{Phys. Rev.} \textbf{\bibinfo{volume}{D67}},
  \bibinfo{pages}{074024} (\bibinfo{year}{2003}).

\bibitem[{\citenamefont{{Madsen}}(2001)}]{2001PhRvL..87q2003M}
\bibinfo{author}{\bibfnamefont{J.}~\bibnamefont{{Madsen}}},
  \bibinfo{journal}{Phys. Rev. Lett.} \textbf{\bibinfo{volume}{87}},
  \bibinfo{pages}{172003} (\bibinfo{year}{2001}).

\bibitem[{\citenamefont{Lugones and Horvath}(2002)}]{PhysRevD.66.074017}
\bibinfo{author}{\bibfnamefont{G.}~\bibnamefont{Lugones}} \bibnamefont{and}
  \bibinfo{author}{\bibfnamefont{J.~E.} \bibnamefont{Horvath}},
  \bibinfo{journal}{Phys. Rev. D} \textbf{\bibinfo{volume}{66}},
  \bibinfo{pages}{074017} (\bibinfo{year}{2002}).

\bibitem[{\citenamefont{{Olesen} and {Madsen}}(1994)}]{1994PhRvD..49.2698O}
\bibinfo{author}{\bibfnamefont{M.~L.} \bibnamefont{{Olesen}}} \bibnamefont{and}
  \bibinfo{author}{\bibfnamefont{J.}~\bibnamefont{{Madsen}}},
  \bibinfo{journal}{\prd} \textbf{\bibinfo{volume}{49}}, \bibinfo{pages}{2698}
  (\bibinfo{year}{1994}).

\bibitem[{\citenamefont{Bombaci et~al.}(2004)\citenamefont{Bombaci, Parenti,
  and Vidana}}]{Bombaci:2004mt}
\bibinfo{author}{\bibfnamefont{I.}~\bibnamefont{Bombaci}},
  \bibinfo{author}{\bibfnamefont{I.}~\bibnamefont{Parenti}}, \bibnamefont{and}
  \bibinfo{author}{\bibfnamefont{I.}~\bibnamefont{Vidana}},
  \bibinfo{journal}{Astrophys. J.} \textbf{\bibinfo{volume}{614}},
  \bibinfo{pages}{314} (\bibinfo{year}{2004}).

\bibitem[{\citenamefont{Vucetich and Horvath}(1998)}]{PhysRevD.57.5959}
\bibinfo{author}{\bibfnamefont{H.}~\bibnamefont{Vucetich}} \bibnamefont{and}
  \bibinfo{author}{\bibfnamefont{J.~E.} \bibnamefont{Horvath}},
  \bibinfo{journal}{Phys. Rev. D} \textbf{\bibinfo{volume}{57}},
  \bibinfo{pages}{5959} (\bibinfo{year}{1998}).

\bibitem[{\citenamefont{{Cheng} and {Usov}}(2006)}]{2006PhRvD..74l7303C}
\bibinfo{author}{\bibfnamefont{K.~S.} \bibnamefont{{Cheng}}} \bibnamefont{and}
  \bibinfo{author}{\bibfnamefont{V.~V.} \bibnamefont{{Usov}}},
  \bibinfo{journal}{\prd} \textbf{\bibinfo{volume}{74}},
  \bibinfo{pages}{127303} (\bibinfo{year}{2006}).

\bibitem[{\citenamefont{{Jaikumar} et~al.}(2006)\citenamefont{{Jaikumar},
  {Reddy}, and {Steiner}}}]{2006PhRvL..96d1101J}
\bibinfo{author}{\bibfnamefont{P.}~\bibnamefont{{Jaikumar}}},
  \bibinfo{author}{\bibfnamefont{S.}~\bibnamefont{{Reddy}}}, \bibnamefont{and}
  \bibinfo{author}{\bibfnamefont{A.~W.} \bibnamefont{{Steiner}}},
  \bibinfo{journal}{Phys. Rev. Lett.} \textbf{\bibinfo{volume}{96}},
  \bibinfo{pages}{041101} (\bibinfo{year}{2006}).

\bibitem[{\citenamefont{{Migdal} et~al.}(1990)\citenamefont{{Migdal},
  {Saperstein}, {Troitsky}, and {Voskresensky}}}]{1990PhR...192..179M}
\bibinfo{author}{\bibfnamefont{A.~B.} \bibnamefont{{Migdal}}},
  \bibinfo{author}{\bibfnamefont{E.~E.} \bibnamefont{{Saperstein}}},
  \bibinfo{author}{\bibfnamefont{M.~A.} \bibnamefont{{Troitsky}}},
  \bibnamefont{and} \bibinfo{author}{\bibfnamefont{D.~N.}
  \bibnamefont{{Voskresensky}}}, \bibinfo{journal}{Physics Reports}
  \textbf{\bibinfo{volume}{192}}, \bibinfo{pages}{179} (\bibinfo{year}{1990}).

\bibitem[{\citenamefont{Yakovlev and Pethick}(2004)}]{Yakovlev:2004iq}
\bibinfo{author}{\bibfnamefont{D.~G.} \bibnamefont{Yakovlev}} \bibnamefont{and}
  \bibinfo{author}{\bibfnamefont{C.~J.} \bibnamefont{Pethick}},
  \bibinfo{journal}{Ann. Rev. Astron. Astrophys.}
  \textbf{\bibinfo{volume}{42}}, \bibinfo{pages}{169} (\bibinfo{year}{2004}).

\end{thebibliography}

\end{document}